\documentclass[preprintnumbers,amsmath,11pt,amssymb,floatfix,superscriptaddress,showpacs,nofootinbib]{revtex4}
\usepackage{graphicx}
\usepackage{epsfig}
\usepackage{bm}
\usepackage{amsfonts}

\input epsf

\begin{document}

\title{\Large Reconstruction of $f(G)$ Gravity with New Agegraphic Dark Energy Model}

\author{Abdul Jawad}
\email{jawadab181@yahoo.com} \affiliation{Department of
Mathematics, University of the Punjab, Quaid-e-Azam Campus,
Lahore-54590, Pakistan.}

\author{Surajit Chattopadhyay}
\email{surajit_2008@yahoo.co.in, surajcha@iucaa.ernet.in}
\affiliation{ Pailan College of Management and Technology, Bengal
Pailan Park, Kolkata-700 104, India.}

\author{Antonio Pasqua}
\email{toto.pasqua@gmail.com} \affiliation{Department of Physics,
University of Trieste, Trieste, Italy.}

\date{\today}

\begin{abstract}
In this work, we consider the reconstruction scenario of new agegraphic dark
energy (NADE) model and $f(G)$ theory of gravity with $G$ representing the Gauss-Bonnet
invariant in the flat FRW spacetime. In this context, we assume a solution of the scale factor in power-law form and study the
correspondence scenario. A new agegraphic $f(G)$ model is
constructed and discussed graphically for the evolution of the
universe. Using this model, we investigate the different eras of the
expanding universe and stability with the help of the equation of state (EoS)
parameter $\omega_{eff}$ and squared speed of sound $v_s^2$, respectively. It is mentioned
here that the reconstructed model represents the quintessence era of
the accelerated expansion of the universe with instability. Moreover,
the statefinder trajectories are studied and we find out that the model is
not capable of reaching the $\Lambda$CDM phase of the universe.

\end{abstract}


\maketitle

\section{Introduction}

The agegraphic DE model is initiated from quantum mechanics through
the uncertainty relation along with the gravitational consequences
in general relativity. This model takes the spacetime and
fluctuations of matter content for the observed DE in the universe.
The original agegraphic DE model was proposed by Cai \cite{cai} to
discuss the expansion of the universe with acceleration. It contains
the age $(T)$ of the universe in the expression of energy density as
$\rho_{\Lambda}=3c^2M_p^2T^{-2}$. The numerical factor $3c^2$ is
used to overcome some uncertaninties and $M_p^2$ represents the
reduced Plank mass. However, this model bears some drawbacks, for
example it does not explain matter dominated era of the universe.
Wei $\&$ Cai \cite{wei} proposed a new model by replacing age of the
universe with the conformal time $(\eta)$ and called it new
agegraphic DE model. This model may naturally solve the coincidence
problem \cite{wei1}. Myung $\&$ Seo \cite{myung} argued that NADE
and holographic DE (HDE) models have same energy densities, the only
difference is that the former has conformal time while latter has
future event horizon. However, there exists some essential
differences between them \cite{wei}.

Another approach to explore the accelerated expansion of the
universe is modified theories of gravities. Between the most studied models of modfied gravity, we mention $f(R)$
gravity \cite{nojiri,nojiri1} where $R$ is the Ricci scalar curvature, $f(T)$
gravity \cite{linder} inherits torsion scalar $T$, $f(G)$ gravity
takes $G$ as Gauss-Bonnet invariant \cite{rastkar}, $f(R,G)$ gravity
and many more. These theories of modified gravity play a vital role, not in
explaining the expanding universe but also many other issues
\cite{nojiri2} in this respect. The $f(G)$ gravity have a de-Sitter
point which is used for the cosmic expansion of the accelerated
universe \cite{nojiri3}. Davis \cite {davis} has found that the model
$f(G)=G^n$ with $n>0$, by keeping $G$ as responsible for DE, is
consistent with solar system tests.

The reconstruction scenario or making correspondences between
different DE models became very appealing in cosmology recently.
Setare $\&$ Saridakis \cite{setare} made correspondence between
$f(G)$ gravity models with HDE model to discuss the cosmic
acceleration of the universe. The NADE model is also discussed in the framework of
Brans-Dicke theory by Liu $\&$ Zhang \cite{liu}. Setare
\cite{setare1} studied the NADE model in $f(R)$ gravity and resulted
that there may exists a phantom-like universe. Jamil $\&$ Saridakis
\cite{jamil} investigated the NADE model in Horava-Lifshitz gravity, which resulted to be
consistent with the observations for the accelerated expanding
universe. The correspondence between $f(G)$ gravity and
entropy-corrected HDE \cite{setare2} leads to the accelerating
universe.

Jawad et al. \cite{jawad} investigated the HDE model in the
framework of $f(G)$ gravity and discussed different phenomenon for
the acceleration of the universe. In this paper, we reconstruct the
NADE model in $f(G)$ gravity by assuming a power-law solution of
the scale factor. We discuss the evolution of the universe
with the help of equation of state (EoS) parameter $\omega_D$ and use squared
speed of sound $v_s^2$ to check the stability of the reconstructed model.
The paper is arranged as follows. In section \textbf{2}, we review
the main features of the $f(G)$ gravity and NADE model. We also construct the
correspondence scenario. In Section \textbf{3}, we provide the
analysis and comparison of reconstructed model. Section \textbf{4}
is devoted to study the statefinder trajectories for the
reconstructed model. The last Section summarizes the obtained
results.

\section{Discussion on $f(G)$ gravity}

The action $S$ for $f(G)$ gravity is given by:
\begin{equation}\label{4}
S=\int
d^4x\sqrt{-g}\left[\frac{1}{2\kappa^{2}}R+f(G)+\mathcal{L}_{m}\right].
\end{equation}
Here $R$ is the Ricci scalar curvature, $\kappa^2=8\pi G_N$ is the coupling
constant with $G_N$ representing the gravitational constant, $g$ is the
determinant of metric tensor $g_{\mu \nu}$, $L_m$ is the matter Lagrangian
and $f(G)$ is an arbitrary differentiable function of $G$ which is generally
defined as:
$G=R^{2}-4R_{\mu\nu}R^{\mu\nu}+R_{\mu\nu\lambda\sigma}R^{\mu\nu\lambda\sigma}$,
where $R_{\mu\nu}$ and $R_{\mu\nu\lambda\sigma}$ are Ricci curvature
tensor and Riemann curvature tensor respectively. By varying the
action $S$ given in  Eq. (\ref{4}) with respect to the metric tensor $g_{\mu \nu}$, the corresponding field
equations are obtained as follow:
\begin{eqnarray}\nonumber
0&=&\frac{1}{2\kappa^2}(-R^{\mu\nu}+\frac{1}{2}g^{\mu
\nu}R)+T^{\mu\nu}+\frac{1}{2}g^{\mu
\nu}f(G)-2f_GRR^{\mu\nu}+4f_GR^{\mu}_{\rho}R^{\nu
\rho}-2f_GR^{\mu\rho\sigma\tau}R^{\nu}_{\rho\sigma\tau}\\\nonumber&-&4f_GR^{\mu\rho\sigma\nu}R_{\rho\sigma}
+2(\nabla^\mu \nabla^\nu f_G)R-2g^{\mu \nu}(\nabla^2
f_G)R-4(\nabla_\rho \nabla^\mu f_G)R^{\nu \rho}-4(\nabla_\rho
\nabla^\nu f_G)R^{\mu
\rho}\\\label{5}&+&4(\nabla^2f_G)R^{\mu}{\nu}+4g^{\mu\nu}(\nabla_\rho
\nabla_\sigma f_G)R^{\rho\sigma}-4(\nabla_\rho \nabla_\sigma
f_G)R^{\mu\rho\nu\sigma},
\end{eqnarray}
where $f_G=f'(G)=\frac{df}{dG}$ is the first derivative with respect to $G$ of the
function $f$ and $T^{\mu \nu}$ represents the energy-momentum tensor of the
perfect fluid.

We here consider a spatially flat (i.e. with curvature parameter $k$ equal to zero) and homogenous FRW spacetime which line element is given by:
\begin{equation}\label{6}
ds^2=-dt^2+a^2(t)(dx^2+dy^2+dz^2),
\end{equation}
with $a(t)$ representing the scale factor which represents the expansion of
the universe and $\left(x,y,z\right)$ the three spatial coordinates. Taking $\kappa^2=1$ and using Eq. (\ref{6}) in
Eq. (\ref{5}), the field equations become:
\begin{eqnarray}\label{7}
0&=&Gf_G-f-24H^3\dot{f}_{G}-6H^2+2\rho_m,\\\label{8}
0&=&8H^2\ddot{f}_G+16H(H^2+\dot{H})\dot{f}_G+f-Gf_G+2(2\dot{H}+3H^2)+2p_m.
\end{eqnarray}
The subscripts $m$ indicates the matter contribution of energy
density and pressure and $H=\frac{\dot{a}}{a}$ is the Hubble
parameter with overdot represening time derivative. The Gauss-Bonnet
invariant $G$ for FRW metric becomes $G=24H^{2}(\dot{H}+H^{2})$. Here
we will incorporate with the pressureless matter, i.e., $p_m=0$
while the energy density of matter$\rho_m$ satisfies the following energy conservation
equation:
\begin{equation}\label{9}
\dot{\rho}_m+3H\rho_m=0.
\end{equation}
This equation has a solution which is given by the following expression:
\begin{equation}\label{10}
\rho_m=\rho_{m0}a^{-3},
\end{equation}
where $\rho_{m0}$ is an arbitrary constant. The field equations given in Eqs.
(\ref{7}) and (\ref{8}) can be rewritten in effective manner as:
\begin{equation}\nonumber
\rho_{eff}=\rho_m+\rho_G=3H^2,\quad p_{eff}=p_G=-(2\dot{H}+3H^2),
\end{equation}
where:
\begin{eqnarray}\label{11}
\rho_{eff}&=&\rho_m+\frac{1}{2}\left[-f
+24H^2\left(H^2+\dot{H}\right)f_G-(24)^2H^4(2\dot{H}^2+H\ddot{H}+4H^2\dot{H})f_{GG}\right],\\\label{12}
p_{eff}&=&\frac{1}{2}\left\{f-24H^2\left(
H^2+\dot{H}\right)f_G+(24)8 H^{2}\left[ 6\dot{H}^{3}
+6H\dot{H}\ddot{H}+24H^{2}\dot{H}^{2}\right.\right.\\\nonumber &+&
\left.\left.6H^{3}\ddot{H}+8H^{4}\dot{H}+H^{2}\dddot{H}
\right]f_{GG}+(24)^28H^4\left(2H^2
+H\ddot{H}+4H^2\dot{H}\right)^2f_{GGG}\right\}.
\end{eqnarray}

\subsection{NADE model}

By replacing the age of the universe $T$ in the energy density of
agegraphic DE model with the conformal time $\eta$ yields the energy density of
NADE model which is given by:
\begin{equation}\label{3}
\rho_{\Lambda}=\frac{3n^2M_p^2}{\eta^2},\quad \eta=\int
\frac{dt}{a(t)},
\end{equation}
We take $M_p^2=1$ just for simplification of the following calculations.

\subsection{Reconstructed dark energy model}

Here we make correspondence between NADE and $f(G)$ model by
equating their energy densities, i.e. $\rho_G=\rho_{\Lambda}$. It yields:
\begin{equation}\label{13}
-f
+24H^2\left(H^2+\dot{H}\right)f_G-(24)^2H^4(2\dot{H}^2+H\ddot{H}+4H^2\dot{H})f_{GG}=\frac{6n^2}{\eta^2}.
\end{equation}
To find out an analytic solution of Eq. (\ref{13}), we assume a power-law form of the
scale factor as follow:
\begin{equation}\label{a}
a(t)=a_0(t_s-t)^n,
\end{equation}
where the constant $a_0$ represents the present day value of the scale factor, $t_s$ is a constant called the finite future singularity
time and $n$ is a real constant. The Hubble parameter $H$, its first and
second derivatives, Gauss-Bonnet invariant and conformal time take
the following forms
\begin{equation}\label{14}
H=\frac{\dot{a}}{a}=-\frac{n}{t_s-t},\quad \dot{H}=\frac{n}{(t_s-t)^2},\quad
\ddot{H}=\frac{2n}{(t_s-t)^3},\nonumber \\
G=\frac{24n^3(1+n)}{(t_s-t)^4},\quad
\eta=\frac{(t_s-t)^{1-n}}{a_0(n-1)}.
\end{equation}

Using Eq.(\ref{14}) in the reconstructed equation, we obtain
\begin{equation}\label{1}
G^2\frac{d^2f}{dG^2}+\frac{(n-1)^2G}{4}\frac{df}{dG}-\frac{(n-1)^2}{4}f
=\frac{3c^2(1 - n)}{a_0}\left[\left(\frac{G}{24 n^3 (n -
1)}\right)^{\frac{n-1}{4}} -\frac{1}{t_s^{n-1}}\right].
\end{equation}
Eq. (\ref{1}) is a second order linear differential equation whose solution is
given by
\begin{eqnarray}\nonumber
f(G)&=&c_1G+c_2G^{-\frac{(n-1)^2}{4}}-\frac{1}{a_0(n^2-6n+5)G}\left[6n(n-10)c^2
t_s^{1-n}G\right.\\\label{2}&+&\left.
4.2^{\frac{11-3n}{4}}3^{\frac{5-n}{4}}n^2(n^2-2n+1)c^2
\left(\frac{G}{n^3(n-1)}\right)^{\frac{n+3}{4}}\right],
\end{eqnarray}
where $c_1$ and $c_2$ are integration constants. This is the
reconstructed NA $f(G)$ DE model which contains a constant
term, a linear term of its argument and two terms depending upon $n$
as well. We restrict here $n>1$ to find out well defined
results.

In the following Section, we discuss the behavior of model
given in Eq. (\ref{2}) along its argument, EoS parameter analysis and stability
of the model with respect to time.

\section{Analysis of reconstructed model}
The first thing we want to do in this Section is to plot the NA $f(G)$ model (\ref{2}) with respect to $G$ to check
its behavior. For this purpose we assume four different values of $n$, i.e.
$n=2.4,2.6,2.8,3$ and fix the other constants involved by taking the values as
$c_1=0.5,c_2=3,c=0.5,~a_0=1$. Figure \textbf{1} shows the behavior
of the model versus $G$. It represents the continuously decreasing
manner of the function for $G<0.6$. The function approaches to the
negative region and becomes approximately constant for $G>1$
incorporating all chosen values of $n$.
\begin{figure}[h] \centering
\epsfig{file=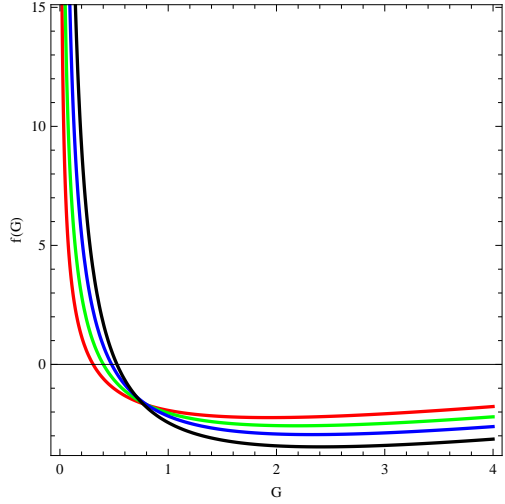,width=.40\linewidth} \caption{Plot of $f(G)$
versus $G$. The curves are characterize as red is for $n=2.4$, green
is for $n=2.6$, blue is for $n=2.8$, black is for $n=3$.}
\end{figure}

The EoS parameter $\omega_{eff}=\frac{p_{eff}}{\rho_{eff}}$ (using
Eqs. (\ref{11}) and (\ref{12}) with corresponding quantities) is
plotted against the cosmic time $t$ for same values of $n$ and constants as for
Figure \textbf{1}. Initially, the behavior of $\omega_{eff}$ indicates a decelerating
phase of the universe as shown in Figure \textbf{2} as
$\omega_{eff}<-\frac{1}{3}$ guarantees for the accelerated universe.
Approximately for $5<t<6$, the graph enters into accelerated region
referred as quintessence era, $-1<\omega_{eff}<-\frac{1}{3}$, for
all values of $n$ considered. The universe remains in this region for $n=2.4$
and $2.6$ as shown by the EoS parameter. The higher values of $n$, i.e.
$n=2.8$ and $3$, indicate the crossing of phantom
divide line $\omega_{eff}=-1$ at $t=6.8$ and $7.6$ respectively and
after this range, the universe faces the phantom era
$\omega_{eff}<-1$. As time elapses, at $t=10$ and $14.5$, the
universe again crosses the phantom divide line and becomes
convergent in quintessence region. Thus, incorporating the first two
values of $n$, the universe behaves as accelerated expanding in
quintessence region and for last two values, the model depicts as
quintom model (crossing of phantom divide line occurs).

\begin{figure}[h]
\begin{minipage}{14pc}
\includegraphics[width=16pc]{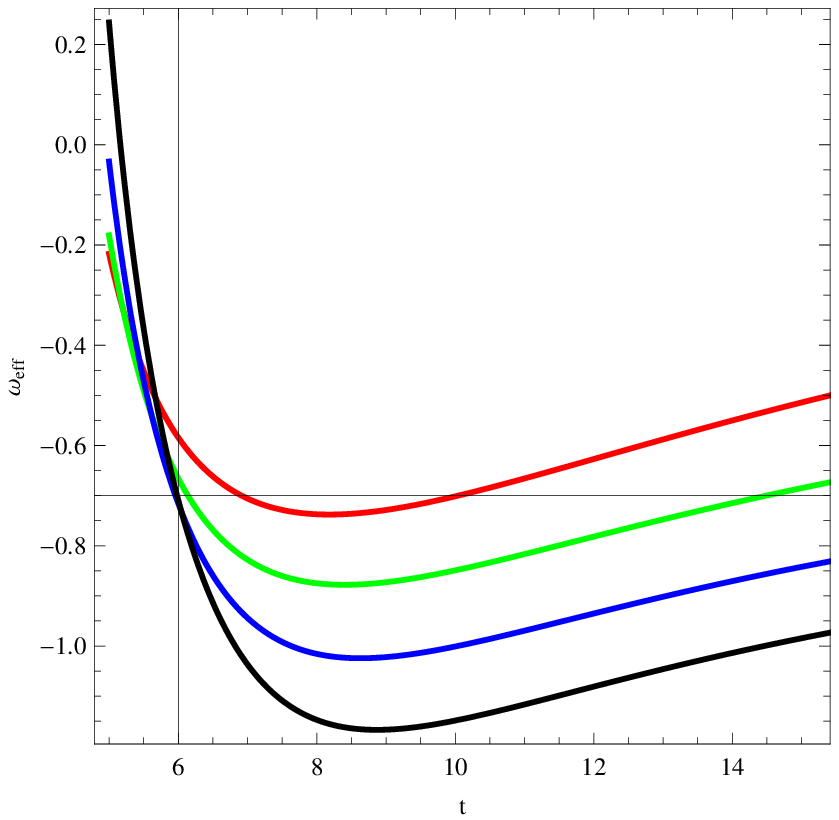}
\caption{\label{label}Plot of $\omega_{eff}$ versus $t$. The curves
are characterize as red is for $n=2.4$, green is for $n=2.6$, blue
is for $n=2.8$, black is for $n=3$.}
\end{minipage}\hspace{3pc}%
\begin{minipage}{14pc}
\includegraphics[width=16pc]{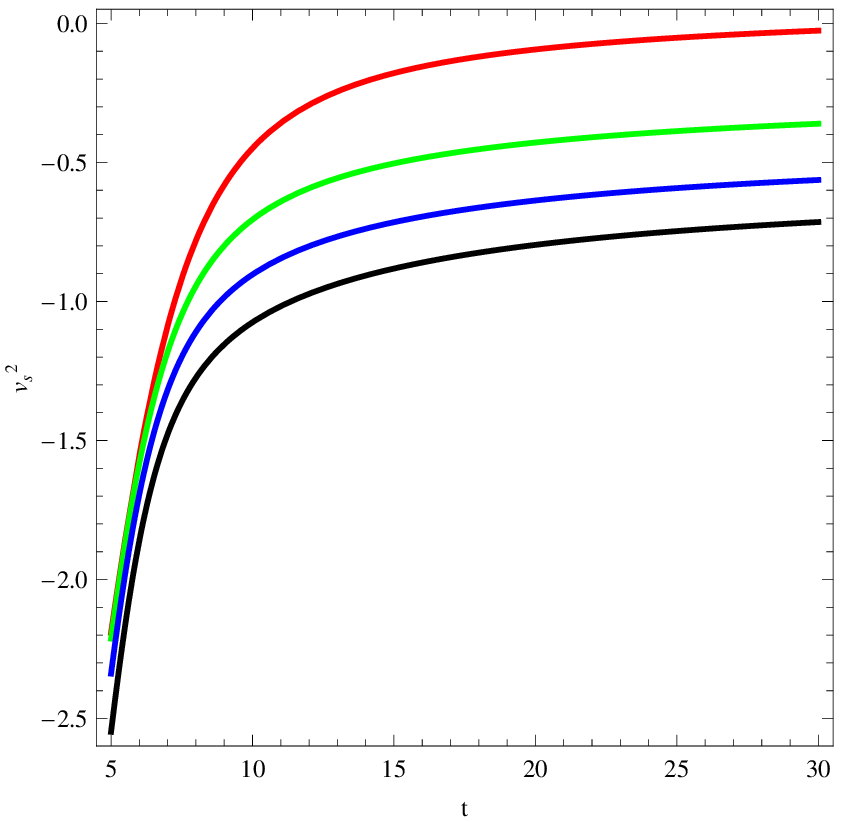}
\caption{\label{label}Plot of $v^2_s$ versus $t$. The curves are
characterize as red is for $n=2.4$, green is for $n=2.6$, blue is
for $n=2.8$, black is for $n=3$.}
\end{minipage}\hspace{3pc}%
\end{figure}
Figure \textbf{3} is a test to analyze the stability of the model by
squared speed sound which defined as the ratio of the time
derivative of total pressure to the time derivative of total energy
density, i.e.:
\begin{equation}\label{15}
v_s^2=\frac{\dot{p}_{eff}}{\dot{\rho}_{eff}}.
\end{equation}
The stability or instability of the model depends upon the sign of
this quantity. Its positive sign corresponds to a stable model while
instability fits for negative values. The stability of many DE
models is discussed using squared speed of sound. Setare
\cite{setare3} studied the interacting HDE model with the
generalized Chaplygin gas model and discussed that both models are
instable. The agegraphic DE model is always negative which shows the
instability of the model explored by Kim et al. \cite{kim}. Jawad et
al. \cite{jawad} checked that the reconstructed HDE $f(G)$ model
always remain instable.

Chattopadhyay and Pasqua \cite{Chattopadhyay} reconstructed the HDE
model in $f(T)$ gravity. They investigated that the squared speed of
sound for reconstructed models bear negative values led to the
classical instable models. Using Eqs.(\ref{11}) and (\ref{12}) for
the NA $f(G)$ model, the $v^2_s$ versus time is plotted taking same
values of $n$ and constants. The graph indicates the instability of
the model for all values of $n$ as squared speed of sound remain
negative during the its evolution. However, we may argue that for
decreasing $n$ it gradually shifts towards zero $v^2_s$ and may
cross it.

\section{Statefinder parameters}

Various candidates for the DE model have been proposed till date and we often face with
the problem of discrimination between them. To get rid of this
problem, Sahni et al. \cite{varun1} introduced the statefinder
$\{r,s\}$ diagnostic pair. This pair has the following form
\cite{varun1,varun2}
\begin{equation}\label{16}
r=\frac{\dddot{a}}{aH^{3}}, \quad
s=\frac{r-1}{3\left(q-\frac{1}{2}\right)}.
\end{equation}
Here $q=-\frac{\ddot{a}}{aH^2}$ is the deceleration parameter and
$H=\frac{\dot{a}}{a}$. Thus the diagnostic pair consists of the
first, second and third derivatives of the scale factor and
indicates geometrically the properties of DE. It has been
discussed for a number of models. Sharif and Jawad \cite{sharif}
discussed the modified HDE in Kaluza-Klein universe and justified
the limit of $\Lambda$CDM model by the statefinder pair.
Panotopoulos \cite{Panotopoulos} discussed this pair for two DE
models. Chakraborty et al. \cite{Chakraborty} discriminated
different DE models in  Kaluza-Klein universe. As demonstrated,
the statefinder can successfully differentiate between a wide
variety of DE models including the Chaplygin gas, the cosmological
constant, braneworld models, quintessence and interacting DE
models \cite{varun1,varun2,Zimdahl}.
\begin{figure}[h]
\includegraphics[width=16pc]{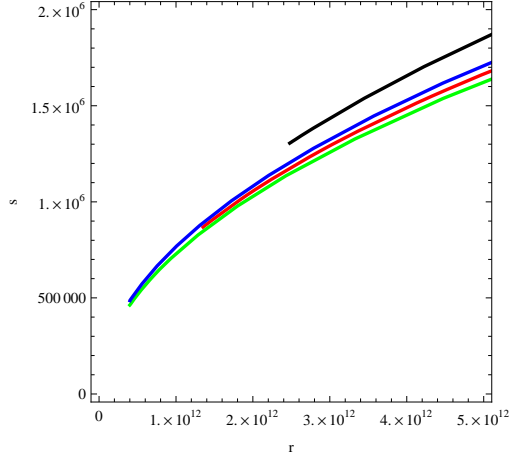}
\caption{\label{label}The $\{r-s\}$ trajectories. The red, green,
blue and black lines correspond to $n=2.9,~2.8,~2.6$ and $2.4$
respectively. Other parameters are same as the previous figures.}
\end{figure}

In the present work, based on the reconstructed energy density
described in the previous Sections, we have created the $\{r-s\}$
trajectories. It is apparent from the $\{r-s\}$ trajectories created
for different choices of parameters stated inside the caption of
Figure \textbf{4} that the trajectories are confined inside the
first quadrant of the $\{r-s\}$ plane. Moreover, it is observed that
although the trajectories are tending towards the fixed point
$\{r=1,s=0\}|_{\Lambda CDM}$, they can not reach it.
Thus, it is inferred that the NA $f(G)$ model is not capable of
reaching the $\Lambda$CDM phase of the universe.

\section{Conclusion}

The search for best fit DE model is very challenging issue in
cosmology recently. In this respect, the reconstruction phenomenon
of different DE models gain a very eminent attention to discuss
accelerated expansion of the universe. The correspondence of HDE,
NADE, their entropy-corrected versions, ghost, scalar field models
within modified gravities give consistent results. In this paper, we
have taken the $f(G)$ gravity framework assuming exact power-law
solution with dust matter. We have considered NADE model in this
framework to explore the cosmic expansion of the universe. The NADE
$f(G)$ model is reconstructed and we have used it for further
prospective. The results are as summarized:

\begin{itemize}
  \item The model contains highly nonlinear terms dependent upon $n$, thus
        we have drawn it for different $n$. The graphical behavior of NADE
        $f(G)$ model has shown the decreasing behavior against $G$.
  \item To characterize the different eras of the universe, we have plot EoS
        parameter versus time $t$ for different values of $n$. For smaller
        values, this parameter shows that the universe stays in the
        quintessence region. The higher values of $n$ represents the
        crossing of phantom divide line twice and the universe lying in
        quintessence state finally. Thus it is interesting to mention here
        that for $n=2.8$ and $3$, the reconstructed NA $f(G)$ model behaves
        like quintom model.
  \item The stability of the model is investigated by squared speed of
        sound. The NADE $f(G)$ model has preserved the instability during
        the evolution of the universe.
  \item The statefinder trajectories are plotted for the reconstructed
        model. It is observed that the NA $f(G)$ model is not
        subject to reach the $\Lambda$CDM model in the underlying scenario.
\end{itemize}

Jawad et al. \cite{jawad} investigated the HDE model in the
framework of $f(G)$ gravity to discuss the acceleration of the
universe and checked the stability of the model with squared speed
of sound. Also, they worked out on energy condition for HDE $f(G)$
model. In the present paper, we found the NADE $f(G)$ model by
making correspondence between NADE model in the underlying gravity.
The reconstructed HDE $f(G)$ model contains a linear term of $G$ and
two nonlinear terms and shows the increasing behavior with respect
to $G$. It shows the increasing behavior graphically. The
reconstructed NADE $f(G)$ model inherits the same type of solution
but highly nonlinear terms dependent upon $n$ with a constant term.
Its graphical behavior represents the decreasing behavior. Both of
the models remain instable during the evolution of the universe.
However the violation of strong energy condition indicates the
accelerated expansion for HDE $f(G)$ model whereas the quintessence
region is observed for acceleration in NADE $f(G)$ gravity with the
help of EoS parameter.

\vspace{0.5cm}

{\bf Acknowledgment}

\vspace{0.5cm}

The first author (AJ) wishes to thank the Higher Education
Commission, Islamabad, Pakistan for its financial support through
the \textit{Indigenous Ph.D. 5000 Fellowship Program Batch-VII}. The
second author (SC) wishes to acknowledge the financial support from
Department of Science and Technology, Govt. of India under Project
Grant no. SR/FTP/PS-167/2011.

\end{document}